\documentclass{emulateapj}

\usepackage{amsmath}
\newcommand{\bo}{\ensuremath{\boldsymbol{B}_0}}
\newcommand{\dm}{\ensuremath{D_{\mu\mu}}}
\newcommand{\const}{\ensuremath{\text{const}}}
\newcommand{\abs}[1]{\ensuremath{\left| #1\right|}}

\newcommand{\Rl}{\ensuremath{R_{\mathrm L}}}

\newcommand{\be}{\begin{equation}}
\newcommand{\ee}{\end{equation}}
\newcommand{\bs}{\begin{subequations}}
\newcommand{\es}{\end{subequations}}
\newcommand{\pa}{\ensuremath{_\parallel}}
\newcommand{\se}{\ensuremath{_\perp}}

\newcommand{\Ph}{\ensuremath{\varPhi}}
\newcommand{\De}{\ensuremath{\Delta}}
\newcommand{\Om}{\ensuremath{\Omega}}

\newcommand{\Ga}{\ensuremath{\Gamma}}
\newcommand{\f}[1]{\ensuremath{\boldsymbol{#1}}}
\newcommand{\m}[1]{\ensuremath{\left\langle #1\right\rangle}}
\newcommand{\pd}[2][]{\ensuremath{\frac{\partial #1}{\partial #2}}}
\newcommand{\dd}[2][]{\ensuremath{\frac{\mathrm{d} #1}{\mathrm{d} #2}}}
\newcommand{\df}{\ensuremath{\mathrm{d}}}

\newcommand{\Artanh}{\mathrm{Artanh}}
\newcommand{\sech}{\mathrm{sech}}

\begin{document}

\shorttitle{PITCH-ANGLE SCATTERING WITH ADIABATIC FOCUSING}
\shortauthors{TAUTZ ET AL.}

\title{Pitch-angle scattering of energetic particles with adiabatic focusing}
\author{R.\,C. Tautz}
\affil{Zentrum f\"ur Astronomie und Astrophysik, Technische Universit\"at Berlin, Hardenbergstra\ss{}e 36, D-10623 Berlin, Germany}
\email{robert.c.tautz@gmail.com}
\author{A. Shalchi}
\affil{Department of Physics and Astronomy, University of Manitoba, Winnipeg, Manitoba R3T 2N2, Canada}
\email{andreasm4@yahoo.com}
\author{A. Dosch}
\affil{\mbox{Center for Space Plasmas and Aeronomic Research, University of Alabama in Huntsville},\\320~Sparkman Drive, Huntsville, AL~35805, USA}
\email{alexanderm.dosch@gmail.com}

%%%%% Abstract
\begin{abstract}
Understanding turbulent transport of charged particles in magnetized plasmas often requires a model for the description of random variations in the particle's pitch angle. The Fokker-Planck coefficient of pitch-angle scattering, which is used to describe scattering parallel to the mean magnetic field, is therefore of central importance. Whereas quasi-linear theory assumes a homogeneous mean magnetic field, such a condition is often not fulfilled, especially for high-energy particles. Here, a new derivation of the quasi-linear approach is given that is based on the unperturbed orbit found for an adiabatically focused mean magnetic field. The results show that, depending on the ratio of the focusing length and the particle's Larmor radius, the Fokker-Planck coefficient is significantly modified but agrees with the classical expression in the limit of a homogeneous mean magnetic field.
\end{abstract}


%%%%% Introduction
\section{Introduction}\label{intro}

The importance of a detailed understanding of charged particle transport had been realized since the early investigations of cosmic-ray acceleration \citep[and references therein]{jok66:qlt}. Due to the typically strong turbulence in astrophysical environments, these diffusion-like processes are significantly different from laboratory and fusion plasmas. Applications range from predicting the so-called ``space weather'' caused by solar energetic particles, over galactic cosmic rays to those rare events with the highest particle energies ever measured \citep[e.\,g.,][]{gle68:mod,for75:ani}.

Any attempt to describe the turbulent transport needs to take into account the structure of the large-scale magnetic field. Especially if a guide magnetic field is only of the same order as a turbulent component and for particles with high energies, the gyration radius may become comparable with the coherence length of the magnetic field. This scale is defined as the typical distance over which the magnetic field appears relatively ordered and corresponds roughly the integral scale of the turbulence \citep{plo11:int}. In the interstellar medium, for example, it is of of the order 1--10\,pc \citep{min96:flu}. Beyond the coherence length, both the velocity field and the magnetic field are turbulent \citep{arm95:lic}, which is not only an observational fact but is also required to confine galactic cosmic rays \citep{jok69:lif} to the Galaxy. Such is in contrast to the usual assumption of a constant background magnetic field, which has been widely used to derive transport parameters \citep[see, e.\,g.,][for an introduction]{rs:rays,sha09:nli}.

In more realistic scenarios, therefore, a reduced (variable) coherence scale needs to be taken into account. Such has been known to introduce additional difficulties, both in the derivation of a turbulence power spectrum \citep{ngs03:tur,sha07:cas} and in the theoretical description of the particle transport \citep[and references therein]{tau12:adf}. It is well-known that non-uniform magnetic fields cause particle drifts \citep{bur10:dri,sch10:nun,tau12:kpa}, which violates the conditions for the applicability of a diffusive description \citep{kot00:foc}. Therefore, the systematic derivation of a transport theory under such circumstances is extremely complicated so that often concentration has focused on singular aspects such as a diverging or converging mean magnetic field, which can be expressed in terms of a so-called ``focusing length'' \citep{roe69:int,kun79:adf}.

Therefore, rather than finding a completely new description, the focusing length has been used to modify the diffusion parameters by including the large-scale geometry of the magnetic field. There are a plethora of practical applications of adiabatic focusing \citep[cf.][]{tau12:adf}, including Solar flares \citep{bie02:obs}, space weather, and in general energetic particle transport ranging from the Earth's magnetosphere \citep{zha06:ept} over the heliospheric termination shock \citep{ler07:foc} to extragalactic radio sources \citep{spa81:adf}.

The analytical determination of the parallel diffusion coefficient is typically based on the Fokker-Planck coefficient of pitch-angle scattering \citep[e.\,g.,][]{has68:sca,ear74:ide}. It seems that currently there is no analytical theory that allows to obtain the parallel diffusion coefficient directly. However, since it has turned out that parallel diffusion is often by at least one order of magnitude stronger than perpendicular diffusion, a one-dimensional treatment based solely on the parallel diffusion is often sufficient. It is therefore necessary to understand in detail the microphysics inherent in the stochastic variations of the pitch-angles due to deflections at magnetic field fluctuations that lead to a description in terms of a Fokker-Planck equation.

However, despite significant progress that has been made towards a reliable description of pitch-angle scattering \citep[e.\,g.,][]{gol76:hom,bie88:pit,jae98:pit,lem09:sto}, an all-encompassing theoretical understanding is still elusive. Here, we show the modifications obtained for the Fokker-Planck coefficient of pitch-angle scattering by introducing a new unperturbed orbit under the assumption of a (constant) adiabatic focusing length for the magnetic field. While the classic quasi-linear result is retained in the limit of a homogeneous magnetic field, it will be shown that in general the Fokker-Planck coefficient---and thus the parallel mean-free path---is strongly altered.

This article is organized as follows: In Sec.~\ref{foc}, the basic principle of adiabatically focused particle transport is introduced and the new unperturbed particle orbit for the quasi-linear framework is derived. Sec.~\ref{fp} explains the derivation of the modified Fokker-Planck coefficient of pitch-angle scattering, and it is shown how the limit of a homogeneous background magnetic field is obtained. In Sec.~\ref{res}, several results are shown for different values of the ratio of the particles' Larmor radius and the focusing length. A comparison with numerical simulation results is given in Sec.~\ref{sim}. Secs.~\ref{disc} and \ref{conc} provide a discussion of the implications and a brief conclusion, respectively.

%%%%%
\section{Adiabatic Focusing}\label{foc}

For a sufficiently tenuous plasma---often realized in astrophysical environments---Coulomb collisions can be neglected and the particle motion is determined by interactions with ambient electromagnetic fields. Such is especially true if the investigation focuses on a population of energetic particles that moves due to turbulence induced by a low-energetic background plasma such as the Solar wind. In that case, the phase-space distribution function, $f$, can be described by a Fokker-Planck equation \citep{roe69:int,luh76:kin,ear76:adf}
\be\label{eq:tra}
\pd[f]t+v\mu\,\pd[f]z+\frac{v}{2L}\left(1-\mu^2\right)\pd[f]\mu=\pd\mu\left(\dm\,\pd[f]\mu\right),
\ee
where terms due to momentum diffusion \citep[cf.][]{lit11:fac} and catastrophic loss processes have been neglected. In Eq.~\eqref{eq:tra}, the pitch-angle cosine has been introduced $\mu=\cos\angle(\f v,\f B)$ as a special coordinate. In addition, the background magnetic field, $\f B$, is assumed to be (approximately) oriented in $z$ direction. The parameter $L$ is known as the magnetic focusing length and will be discussed in the following subsection.

By averaging over the pitch-angle cosine, the Fokker-Planck coefficients are replaced by the diffusion tensor and the equation for $f$ becomes the transport equation \citep[see, e.\,g.,][]{par65:pas,zan13:tra}. Analytically, the basic derivation from the Vlasov-Maxwell equations requires that the magnetic turbulence is stationary, homogeneous, and has a weak magnitude compared to the mean magnetic field \citep[see, e.\,g.,][for an introduction]{rs:rays,sha09:nli}.

\subsection{Focused Magnetic Field}

In the case of a spatially variable mean magnetic field, the effect on the particle distribution can be quantified by introducing the typical length scale of the converging or diverging field. Due to the condition that $\nabla\cdot\f B=0$, both the orientation and the field strength will vary. Physically, the value of the associated parameter $L$ describes the effect of particle trapping and or adiabatic acceleration due to converging and diverging magnetic field lines, respectively. Based on the seminal work by \citet{roe69:int}, the focusing length is introduced as
\be\label{eq:L}
L^{-1}=\nabla\cdot\left(\frac{\f B}{B}\right)=-\frac{1}{B}\,\pd[B]s
\ee
where $s$ denotes the coordinate along the magnetic field and where $B=|\f B|$ is the total field strength. To simplify the investigation, the assumption will be made in what follows that $s$ can be replaced by the Cartesian coordinate $z$ \citep[see Appendix~A of][for a discussion]{tau12:adf}. In addition, a constant focusing length will be chosen. This is in agreement with earlier analytical investigations \citep{ear76:adf,kun79:adf}, where $L=\const$ was assumed on spatial scales that are large compared to the parallel mean free path of the particles. In combination with the second assumption this leads to an exponential behavior of the mean magnetic field as
\bs\label{eq:Bxyz}
\be\label{eq:Bz}
B_z=B_0\,e^{-z/L}.
\ee

For most applications, the magnetic field is chosen to be axisymmetric. This is valid even for applications requiring a fully anisotropic diffusion tensor, such as the Solar wind. Here, the Parker-spiral geometry of the mean magnetic field and the Solar-wind velocity, which induces the turbulent fields, forms a variable angle does not allow for a two-dimensional description \citep[e.\,g.,][and references therein]{eff12:ani}. From Eq.~\eqref{eq:Bz}, the other two magnetic field components are then immediately obtained as
\begin{eqnarray}
B_x&=&B_0\,\frac{x}{2L}\,e^{-z/L}\\
B_y&=&B_0\,\frac{y}{2L}\,e^{-z/L}.
\end{eqnarray}
\es
due to the condition that the divergence of the magnetic field be zero \citep[cf.][]{tau12:adf}.

It should be noted that the procedure can be applied only to relatively smooth configurations \citep[e.\,g.,][]{heh12:mfp}, as opposed to arbitrarily oriented magnetic fields with a twisted structure as obtained, for instance, in magneto-hydrodynamical simulations of the interstellar medium \citep[e.\,g.,][]{avi05:mag}. However, the approach described here should remain valid as long as the curvature radius of the $x$ and $y$ components of the magnetic field are large compared with that of the $z$ component of the field \citep[cf.][]{lit11:fac}. This is ensured by the exponential $z$ dependence as opposed to the linear dependence on $x$ and $y$. The general case has been briefly discussed in Appendix~A of \citet{tau12:adf}.

\subsection{Undisturbed trajectory}\label{undis}

The following derivation is based on the geometry of the ``magnetic bottle'' as detailed, e.\,g., in \citet{che06:pla}. The basic assumption is that the magnetic field is predominantly oriented along a coordinate axis, which, following the convention in transport theory, is the $z$ direction.

As shown in Appendix~\ref{bottle}, the force acting on a charged particle has two components, causing (i) the usual gyro-motion around the magnetic field and (ii) an additional acceleration along the $z$ direction. That force can be expressed as
\be\label{eq:Fz}
F_z=\gamma m\dot v\pa\simeq-\frac{\gamma mv\se^2}{2B}\,\pd[B_z]z,
\ee
where use has been made of the fact that, for a static magnetic field, $\abs{\f v}=\const$ and hence $\gamma=\const$. In Sec.~\ref{fp}, the case of magnetostatic turbulence will be considered, in which case the assumption of a constant particle speed remains true.

%%%%%%%%%%%%%%%%%%%%%%%%%%%%%%%%%%______
\begin{figure}[t]
\centering
\includegraphics[width=\linewidth]{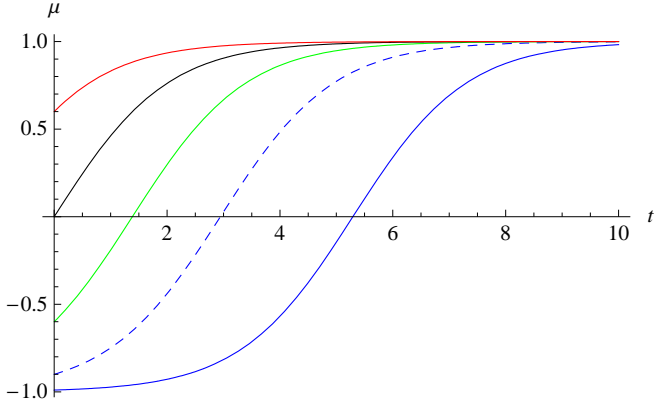}
\caption{(Color online) Pitch-angle $\mu$ as a function of time $t$. Initial pitch-angle cosines are $\mu_0=-0.9$ (solid blue line), $\mu_0=-0.9$ (dashed blue line), $\mu_0=-0.6$ (green line), $\mu_0=0$ (black line), and $\mu_0=0.6$ (red line).}
\label{ab:mu_of_t}
\end{figure}
%%%%%%%%%%%%%%%%%%%%%%%%%%%%%%%%%%^^^^^^

In the limit that the magnetic field components perpendicular to the $z$ axis are small, i.\,e., $B_x\ll B_z$ and $B_y\ll B_z$, the focusing length from Eq.~\eqref{eq:L} is approximately $L=-B_z/(\partial B_z/z)$. Inserting this into the parallel force component from Eq.~\eqref{eq:Fz}, one has
\be
\dot v\pa\simeq\frac{v\se^2}{2L},
\ee
which, by using the fact that the total velocity $v^2=v\pa^2+v\se^2$ is conserved, can be rewritten as a differential equation for the parallel velocity component as
\be\label{eq:dotvpa}
\dot v\pa\simeq\frac{v^2}{2L}-\frac{v\pa^2}{2L}.
\ee
With $v\pa(0)=v\mu_0$, this differential equation can be solved in closed analytical form, yielding
\bs\label{eq:vpa1}
\begin{eqnarray}
v\pa(t)&=&v\,\tanh\left(\frac{vt}{2L}+\Artanh\,\mu_0\right) \label{eq:vpa1a}\\
&=&v-\frac{2v\left(1-\mu_0\right)}{1-\mu_0+\left(1+\mu_0\right)e^{vt/L}},
\end{eqnarray}
\es
where the first form is analytically more appealing, while the second form is more suitable for numerical evaluations. If $\mu_0=0$, the particle is initially at a turning point (i.\,e., the pitch angle is $90^\circ$).

%%%%%%%%%%%%%%%%%%%%%%%%%%%%%%%%%%______
\begin{figure}[t]
\centering
\includegraphics[width=\linewidth]{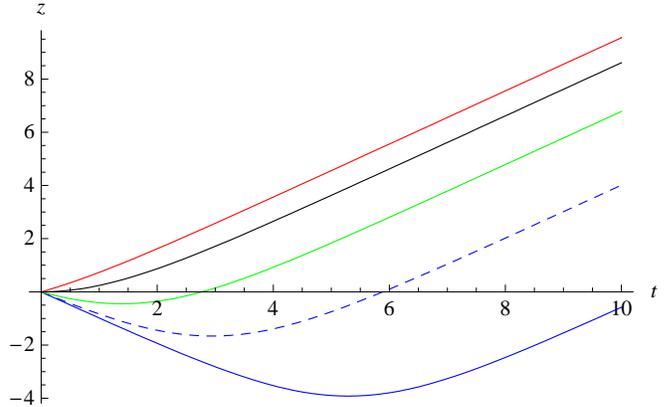}
\caption{(Color online) Parallel coordinate $z$ as a function of time $t$. Initial pitch-angle cosines are $\mu_0=-0.9$ (solid blue line), $\mu_0=-0.9$ (dashed blue line), $\mu_0=-0.6$ (green line), $\mu_0=0$ (black line), and $\mu_0=0.6$ (red line).}
\label{ab:z_of_t}
\end{figure}
%%%%%%%%%%%%%%%%%%%%%%%%%%%%%%%%%%^^^^^^

Due to $v^2=v\pa^2+v\se^2=\const$, one has
\bs\label{eq:vse}
\begin{eqnarray}
v\se&=&\sqrt{v^2-v\pa^2}=v\;\sech\left(\frac{vt}{2L}+\Artanh\,\mu_0\right)\\
&=&v\sqrt{1-\left[1-\frac{2\left(1-\mu_0\right)}{1-\mu_0+\left(1+\mu_0\right)e^{vt/L}}\right]^2},
\end{eqnarray}
\es
which is plausible for particles traveling toward weaker magnetic fields: even if initially $v\pa=0$ (and so $\mu_0=0$), a particle will eventually move parallel to an (exponentially) decreasing magnetic field. In the limit of weak focusing, i.\,e., $L\to\infty$, Eq.~\eqref{eq:vse} gives $v\se\approx v\sqrt{1-\mu_0^2}$ as required.

Furthermore, the parallel coordinate, $z(t)$, can be obtained by integrating Eq.~\eqref{eq:vpa1} over time, which yields
\bs\label{eq:z}
\begin{eqnarray}
z(t)&=&2L\;\ln\left[\cosh\left(\frac{vt}{2L}+\Artanh\,\mu_0\right)\right]\nonumber\\
&+&L\,\ln\left(1-\mu_0^2\right)+z_0\\
&=&2L\left[\frac{1}{2}\left(1-\mu_0+\left(1+\mu_0\right)e^{vt/L}\right)\right]\nonumber\\
&-&vt+z_0,
\end{eqnarray}
\es
with the integration constant chosen so that so that $z(0)=z_0$. In Figs.~\ref{ab:mu_of_t} and \ref{ab:z_of_t}, $\mu(t)$ and $z(t)$ are shown for different initial pitch-angles, respectively. Particles initially moving toward stronger magnetic fields at some point turn around and move toward weaker magnetic fields, just as expected.\\

By combining the results and using the Cartesian representation of the Lorentz force, the time derivative of the parallel velocity components reads
\be
\dot v\pa(t)=\frac{q}{mc}\left[v_x\left(B_y+\delta B_y\right)-v_y\left(B_x+\delta B_x\right)\right],
\ee
with $v_x$ and $v_y$ from Eqs.~\eqref{eq:vxy} and $v\se$ from Eq.~\eqref{eq:vse}.

%%%%%
\section{Fokker-Planck Coefficient}\label{fp}

The general definition of the Fokker-Planck coefficient for pitch-angle scattering reads \citep{sha09:nli}
\bs
\begin{eqnarray}
\dm&=&\int_0^\infty\df t\;\m{\dot\mu(t)\,\dot\mu(0)} \label{eq:dm_int}\\
&=&\frac{1}{2}\,\dd t\m{\left(\De\mu(t)\right)^2}\approx\frac{1}{2t}\,\m{\left(\De\mu(t)\right)^2}, \label{eq:dm_tgk}
\end{eqnarray}
\es
with $\De\mu(t)=\mu(t)-\mu(0)$, which form can be derived systematically from the relativistic Vlasov equation \citep[e.\,g.,][]{rs:rays,sha11:tgk}.

Note also that, according to Eq.~\eqref{eq:dm_tgk}, the Fokker-Planck coefficient is usually taken to be a function of the pitch-angle cosine, $\mu$. However, a perturbation theory for the description of particle scattering always requires the specification of a particle trajectory, which connects $\mu$ to the time, $t$, with initially $\mu(t=0)=\mu_0$. For a homogeneous background magnetic field, $\mu=\mu_0\;\forall t$ so that the functional dependence on $\mu$ is seemingly retained even though, in Eq.~\eqref{eq:dm_tgk}, one integrates over time. However, if $\mu$ is a non-trivial function of time, then the functional dependence of $\dm$ on $\mu$ is lost during the time integration, unless \emph{both} the dependence on both $\mu$ and the time are taken into account. Such would require $\dm$ in Eq.~\eqref{eq:dm_int} to be a function of $t$, where $t$ is the upper limit of the integral.

In what follows, the two contributions to the Fokker-Planck coefficient will be derived that arise from: (i) the large-scale geometry that causes changes in the particle motion; and (ii) the traditional turbulent scattering.

\subsection{Large-scale contribution}

The large-scale contribution to the Fokker-Planck can be obtained by using $\dot\mu=\dot v\pa/v$ as given by Eq.~\eqref{eq:dotvpa}. In addition, one requires $\dot\mu(0)=v(1-\mu_0^2)/(2L)$ and, for the insertion of $\dot\mu(t)\dot\mu(0)$ into Eq.~\eqref{eq:vpa1a}, the integral over the squared pitch angle,
\be
\int_0^\infty\df t\,\left(1-\mu(t)^2\right)=\frac{2L}{v}\left(1-\mu_0\right).
\ee
The resulting Fokker-Planck coefficient reads
\be\label{eq:dm0}
\dm^{(0)}(\mu_0)=\frac{v}{2L}\left(1-\mu_0^2\right)\left(1-\mu_0\right),
\ee
where the index~$(0)$ indicates that Eq.~\eqref{eq:dm0} describes the systematic changes in the particle motion due to the curvature of the mean magnetic field. The resulting effect is strongest for particles with negative initial pitch angles as they are reflected due to the magnetic mirror effect (cf. Fig.~\ref{ab:mu_of_t}).

%%%%%%%%%%%%%%%%%%%%%%%%%%%%%%%%%%______
\begin{figure}[t]
\centering
\includegraphics[width=1.08\linewidth]{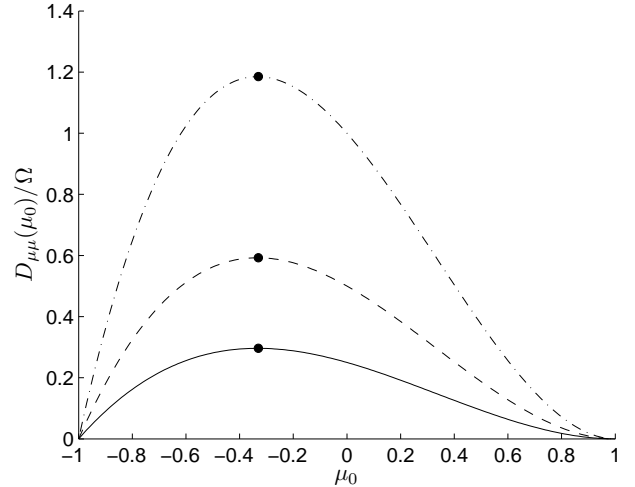}
\caption{Large-scale contribution to the Fokker-Planck coefficient, $\dm^{(0)}(\mu_0)$, as given in Eq.~\eqref{eq:dm0}. Shown are the cases $L=2\ell_0$ (solid line), $L=\ell_0$ (dashed line), and $L=\ell_0/2$ (dotdashed line), with $R=1$ always. The black dots mark the maximum of $\dm^{(0)}(\mu_0)$, which is always located at $\mu_0=-0.3$.}
\label{ab:dm_unperturbed}
\end{figure}
%%%%%%%%%%%%%%%%%%%%%%%%%%%%%%%%%%^^^^^^

In Fig.~\ref{ab:dm_unperturbed}, the large-scale contribution to the Fokker-Planck coefficient is shown for different values of the focusing length, $L$. By using the connection with the parallel scattering mean-free path, a mean-free path equivalent can be obtained as \citep{ear74:ide}
\be\label{eq:la0}
\lambda\pa^{(0)}=\frac{3v}{8}\int_{-1}^1\df\mu_0\;\frac{\left(1-\mu_0^2\right)^2}{\dm^{(0)}(\mu_0)}=\frac{3L}{2}.
\ee
This result merely illustrates that a curved magnetic field behaves as though the particles were scattered in a classic sense due to a turbulent magnetic field. In Sec.~\ref{res}, therefore, only the contribution due to the turbulent magnetic field, as derived in the following subsection, is taken into account. Note, however, that a more detailed derivation reveals that the relation between the Fokker-Planck coefficient and the parallel mean-free path is modified, as shown, e.\,g., by \citet[his Eq.~(33)]{sha11:adf}. In the case of Eq.~\eqref{eq:la0}, however, such results only in the reduction of the parallel mean-free path by a constant factor of $\sim0.8$ so that the general behavior remains unchanged.

It has been shown before analytically and numerically \citep[see, e.\,g.,][and references in both papers]{sha13:ana,dan13:fpe} that the standard relation between the Fokker-Planck coefficient and the parallel mean-free path as shown in Eq.~\eqref{eq:la0} is no longer valid if there is a non-vanishing focusing term. Therefore, focusing changes fundamental quantities and relations in transport theories in two ways.

\begin{enumerate}
\item The relation between $\dm$ and $\lambda\pa$ is much more complicated. This effect is well-know and was investigated previously in several papers \citep{sha13:ana,dan13:fpe}. In this case the parallel mean free path with focusing depends on the mean-free path without focusing as well as on the focusing length. In this approach the focusing-free Fokker-Planck coefficient is used.
\item The Fokker-Planck coefficient itself can be different if there is focusing. This case was not investigated in previous work and is subject of the present article.
\end{enumerate}

\subsection{Turbulent contribution}

The second contribution to $\dm$ can be obtained by making use of $\dot v\pa=v\dot\mu$. Accordingly, the time derivative of the pitch-angle cosine can be expressed by the parallel velocity component, which has two contributions
\be
\dot\mu=\frac{q}{mcv}\left[v_x\left(B_y+\delta B_y\right)-v_y\left(B_x+\delta B_x\right)\right],
\ee
where the magnetic field components from Eqs.~\eqref{eq:Bxyz} have to be inserted.

Now for the pitch-angle Fokker-Planck coefficient, $\dm$, one needs
\begin{eqnarray}
\dot\mu(t)\dot\mu(0)&=&\left(\frac{q}{mcv}\right)^2\biggl[
  v_x(t)v_x(0)\left(B_y+\delta B_y\right)_t\left(B_y+\delta B_y\right)_0\nonumber\\
&-&v_x(t)v_y(0)\left(B_y+\delta B_y\right)_t\left(B_x+\delta B_x\right)_0\nonumber\\[8pt]
&-&v_y(t)v_x(0)\left(B_x+\delta B_x\right)_t\left(B_y+\delta B_y\right)_0\nonumber\\[3pt]
&+&v_y(t)v_y(0)\left(B_x+\delta B_x\right)_t\left(B_x+\delta B_x\right)_0\biggr], \label{eq:DmFull}
\end{eqnarray}
where the subscripts $t$ and $0$ mean ``to be evaluated at time $t$ and time $t=0$'', respectively.

To proceed, several assumptions have to be made that allow the analytical tractability of Eq.~\eqref{eq:DmFull}:
\begin{itemize}
\item $x(0)=y(0)=0$ so that $B_x(0)=B_y(0)=0$\\[-3ex]
\item $\m{\delta B_x(t)\,\delta B_y(0)}=\m{\delta B_y(t)\,\delta B_x(0)}=0$\\[-3ex]
\item $\m{\delta B_x(t)\,\delta B_x(0)}=\m{\delta B_y(t)\,\delta B_y(0)}=\mathsf R_{xx}$
\end{itemize}
with $\mathsf R_{ij}$ the magnetic two-point, two-time correlation tensor in position space.
Note that, without the first condition, several additional terms would appear in the following equations, i.\,e., after applying the correlation operator.

For the correlation of the pitch-angle time derivative it follows that
\begin{eqnarray}
\m{\dot\mu(t)\,\dot\mu(0)}&=&\left(\frac{q}{mcv}\right)^2\m{\delta B_x(t)\,\delta B_x(0)}\nonumber\\
&\times&\left(v_x(t)v_x(0)+v_y(t)v_y(0)\right).
\end{eqnarray}
In the quasi-linear theory \citep{jok66:qlt}, the unperturbed orbit is used for the calculation of particle scattering. In contrast to a homogeneous background magnetic field, in which case said unperturbed motion corresponds to $\mu=\mu_0$, now the more complex velocity components derived in Sec.~\ref{undis} has to be used. Therefore, by inserting Eq.~\eqref{eq:vse} in Eqs.~\eqref{eq:vxy} one has
\bs
\begin{eqnarray}
v_x(t)v_x(0)&=&v^2\sqrt{1-\mu_0^2}\,\sech\left(\frac{vt}{2L}+\Artanh\,\mu_0\right)\nonumber\\
&\times&\cos\left(\Ph_0-\Om t\right)\cos\Ph_0\\
v_y(t)v_y(0)&=&v^2\sqrt{1-\mu_0^2}\,\sech\left(\frac{vt}{2L}+\Artanh\,\mu_0\right)\nonumber\\
&\times&\sin\left(\Ph_0-\Om t\right)\sin\Ph_0.
\end{eqnarray}
\es
The next step is to average over the gyro phase, $\Ph_0$, which leads to
\begin{eqnarray}
&&\frac{1}{2\pi}\int_0^{2\pi}\df\phi\bigl[\cos\left(\Ph_0-\Om t\right)\cos\Ph_0\nonumber\\
&&+\sin\left(\Ph_0-\Om t\right)\sin\Ph_0\bigr]=\cos(\Om t).
\end{eqnarray}

A Fourier transformation allows one to introduce the magnetic correlation tensor in wavenumber space, $\mathsf P_{ij}(\f k)$, which, in the magnetostatic limit, reads
\be
\mathsf P_{ij}(\f k)=\m{\delta B_i(\f k)\,\delta B_j^\star(\f k)},
\ee
with $i,j\in\{x,y,z\}$. Then one has
\begin{eqnarray}
\m{\dot\mu(t)\,\dot\mu(0)}&=&\frac{\Om^2}{B_0^2}\,\sqrt{1-\mu_0^2}\;\cos(\Om t)\nonumber\\
&\times&\int\df^3k\,\m{\delta B_x(\f k)\,\delta B_x^\star(\f k)\,e^{i\f k\cdot\f r}}\nonumber\\
&\times&\sech\left(\frac{vt}{2L}+\Artanh\,\mu_0\right). \label{eq:fourier}
\end{eqnarray}
According to quasi-linear theory, the unperturbed trajectory is again inserted on the right-hand side. The Fourier factor is therefore well-defined and can be approximated via $\langle e^{i\f k\cdot\f r}\rangle =e^{ik\pa\,z(t)}$, i.\,e., through the parallel motion \citep[cf.][]{sha05:soq}.

For simplification, the case of magnetostatic slab geometry is considered, which is characterized as $\delta\f B(\f r,t)=\delta\f B(z)$ so that the Fourier modes are aligned with the predominant direction of the mean magnetic field. The correlation tensor component is then given through
\be
\mathsf P_{xx}(\f k)=\m{\delta B_x(\f k)\,\delta B_x^\star(\f k)}=G(k\pa)\,\frac{\delta(k\se)}{k\se},
\ee
with the turbulence power spectrum \citep{sha09:flr}
\begin{eqnarray}
G(k\pa)&=&\frac{D(s,q)}{2\pi}\;\delta B^2\ell_0\;\frac{\abs{\ell_0k}^q}{\left(1+\ell_0^2k^2\right)^{(s+q)/2}}\nonumber\\
&\equiv&\frac{D(s,q)}{2\pi}\;\delta B^2\ell_0\;g(k\pa),
\end{eqnarray}
where $s=5/3$ and $q$ are the inertial and energy range spectral indices, respectively. The normalization function, $D(s,q)$, is obtained from the condition that the integral of the trace of the correlation tensor has to be equal to the total turbulence strength so that
\be
D(s,q)=\frac{\Ga\bigl((s+q)/2\bigr)}{2\Ga\bigl((s-1)/2\bigr)\,\Ga\bigl((q+1)/2\bigr)}.
\ee
For the Fokker-Planck coefficient it follows that
\begin{eqnarray}
\dm&=&\frac{\Om^2}{\abs{\f B}^2}\sqrt{1-\mu_0^2}\int\df^3k\;G(k\pa)\,\frac{\delta(k\se)}{k\se}\nonumber\\
&\times&\int_0^\infty\df t\;\sech\left(\frac{vt}{2L}+\Artanh\,\mu_0\right)\nonumber\\
&\times&\left[\cos\left(k\pa z(t)+\Om t\right)+\cos\left(k\pa z(t)-\Om t\right)\right].
\end{eqnarray}

Finally, use normalized variables $\tau=\Om t$, $R=v/(\Om\ell_0)$ and normalize $L$ as $\tilde L=L/\ell_0$, too. The resulting form for the Fokker-Planck coefficient $\dm$ reads
\begin{eqnarray}
\frac{\dm}{\Om}&=&2D(s,q)\left(\frac{\delta B}{\abs{\f B}}\right)^2\sqrt{1-\mu_0^2}\int_0^\infty\df x\;g(x)\nonumber\\
&\times&\int_0^\infty\df\tau\;\sech\left(\frac{R\tau}{2\tilde L}+\Artanh\,\mu_0\right)\nonumber\\[5pt]
&\times&\left[\cos\bigl(x\,z(\tau)+\tau\bigr)+\cos\bigl(x\,z(\tau)-\tau\bigr)\right]. \label{eq:DmFin}
\end{eqnarray}
Therefore, $\dm$ now depends on the initial pitch-angle cosine, $\mu_0$, since $\mu$ has been inserted as a function of $t$.

\subsection{Limit of Weak Focusing}

First, in the formal limit of $L\to\infty$, one has
\bs
\begin{eqnarray}
\lim_{L\to\infty}\sech\left(\frac{vt}{2L}+\Artanh\,\mu_0\right)&=&\sqrt{1-\mu_0^2}\\
\lim_{L\to\infty}\tanh\left(\frac{vt}{2L}+\Artanh\,\mu_0\right)&=&\mu_0\equiv\mu
\end{eqnarray}
\es
so that the parallel coordinate yields
\be
z(t)=v\mu_0t=v\mu t,
\ee
which corresponds to the unperturbed motion without adiabatic focusing as used in the classic quasi-linear theory. Note that, alternatively, the latter expression for the parallel coordinate, $z(t)$, can be obtained by taking the limit $L\to\infty$ in Eq.~\eqref{eq:z}.
In that case, the Fokker-Planck coefficient is reduced to the well-known form \citep[e.\,g.,][]{sha05:soq}
\bs
\begin{eqnarray}
\dm&=&2D(s,q)\,\frac{\Om^2}{\abs{\f B}^2}\left(1-\mu_0^2\right)\int\df^3k\;G(k\pa)\,\frac{\delta(k\se)}{k\se}\nonumber\\
&\times&\int_0^\infty\df t\left[\cos\left((k\pa v\mu+\Om)t\right)\right.\nonumber\\
&+&\left.\cos\left((k\pa v\mu-\Om)t\right)\right]\\
&=&D(s,q)\,\frac{\pi}{v\abs\mu}\left(\frac{\delta B}{\abs{\f B}}\right)^2\left(1-\mu^2\right)g\!\left(\frac{1}{R\abs\mu}\right). \label{eq:dmu_slab}
\end{eqnarray}
\es
The validity of Eq.~\eqref{eq:dmu_slab} can be tested, for instance, using numerical test-particle simulations. There, the trajectories of a large number of test particles are integrated so that the individual pitch angles can be recorded, thus allowing for a direct evaluation of the Fokker-Planck coefficient via a suitable averaging process.
As shown in \citet{qin09:dmm,tau13:pi1}, the functional dependence of the pitch-angle Fokker-Planck coefficient on $\mu$ can indeed be reproduced---except for pitch angles close to $90^\circ$, i.\,e., at $\mu$ close to zero.

Second, for weakly focused magnetic fields, in which case $L$ is large compared to the Larmor radius, $\Rl=\gamma v_\phi/\Om$, a series expansion of the hyperbolic function entering the parallel velocity components yields
\be
\tanh\left(\frac{vt}{2L}+\Artanh\,\mu_0\right)\approx\mu_0+\frac{vt}{2L}\left(1-\mu_0^2\right)+\mathcal O(L^{-2}),
\ee
thereby providing the first correction to the parallel coordinate as
\be
z(t)\approx v\mu_0t+\frac{v^2t^2}{2L}\left(1-\mu_0^2\right).
\ee
Note that, for the sech function in \dm, a Taylor expansion is \emph{not} possible because the magnitude of $vt/L$ depends on time. But when $L$ is large, the additional increase of $z$ with $t^2$ causes a more rapid oscillation of the cosine factors in \dm, thereby allowing one to use the formal limit of infinitely large $L$ in the sech function. The first correction to the Fokker-Planck coefficient is thereby provided through
\begin{eqnarray}
\dm&=&\frac{\Om^2}{\abs{\f B}^2}\left(1-\mu_0^2\right)\int\df^3k\;G(k\pa)\,\frac{\delta(k\se)}{k\se}\nonumber\\
&\times&\int_0^\infty\df t\left[\cos\left((k\pa v\mu_0+\Om)t+\frac{v^2t^2}{2L}(1-\mu_0^2)\right)\right.\nonumber\\
&+&\left.\cos\left((k\pa v\mu-\Om)t+\frac{v^2t^2}{2L}(1-\mu_0^2)\right)\right]\nonumber\\
&+&\mathcal O(L^{-2}),
\end{eqnarray}
which, on $L\to\infty$, reduces to the well-known quasi-linear limit.

%%%%%
\section{Results}\label{res}

Already from the last term in Eq.~\eqref{eq:DmFin} the obvious observation can be made that $\dm$ is \emph{not} symmetric in $\mu_0$ in contrast to most derivations for homogeneous background magnetic fields. However, such is plausible since the cases of particles initially moving toward stronger magnetic fields ($\mu_0<0$) or weaker magnetic fields ($\mu>0$) are fundamentally different. As shown in Fig.~\ref{ab:mu_of_t}, all particles will eventually attain positive values for their pitch-angles.

\begin{figure}[t]
\centering
\includegraphics[width=1.08\linewidth]{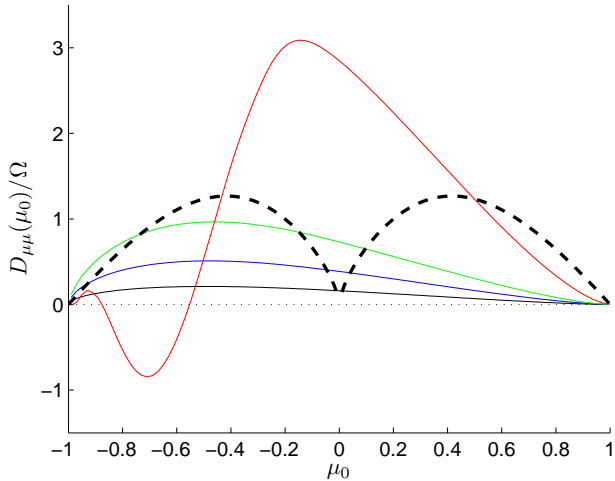}
\caption{(Color online) Fokker-Planck coefficient, \dm, as a function of the initial pitch-angle cosine, $\mu_0$ for small values of the focusing length, $L$. For the normalized rigidity, a value $R=1$ is chosen. In particular, $L=0.02\,\ell_0$ (thin black line), $L=0.05\,\ell_0$ (blue line), $L=0.1\,\ell_0$ (green line), and $L=\ell_0$ (red line) are shown. In addition, the result for $L\to\infty$ is shown as the thick dashed line.}
\label{ab:dm_small}
\end{figure}

\begin{figure}[t]
\centering
\includegraphics[width=1.08\linewidth]{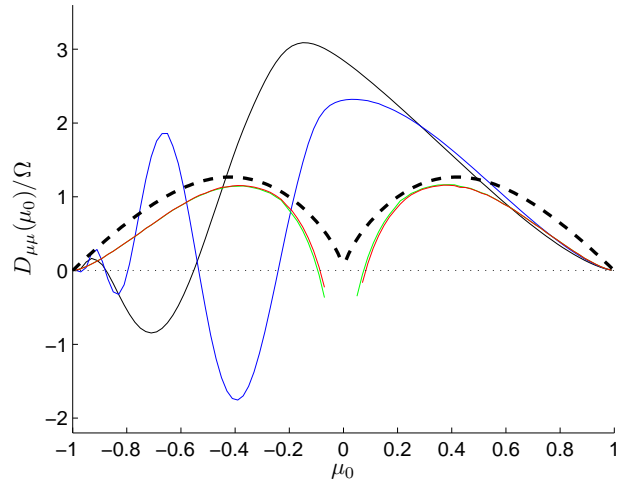}
\caption{(Color online) Fokker-Planck coefficient, \dm, as a function of the initial pitch-angle cosine, $\mu_0$ for large values of the focusing length, $L$. For the normalized rigidity, a value $R=1$ is chosen. In particular, $L=\ell_0$ (thin black line), $L=2\,\ell_0$ (blue line), $L=200\,\ell_0$ (green line), and $L=1000\,\ell_0$ (red line) are shown. In addition, the result for $L\to\infty$ is shown as the thick dashed line.}
\label{ab:dm_large}
\end{figure}

The result is shown in Figs.~\ref{ab:dm_small} and \ref{ab:dm_large} for different values of the focusing length, $L$, as the initial pitch-angle cosine is varied in the interval $-1\leqslant\mu_0\leqslant1$. The two figures reveal the following features:

\begin{itemize}
\item For $L$ smaller than the Larmor radius, i.\,e., $L\lesssim\Rl$, only one maximum is exhibited, which is asymmetrically located at negative $\mu_0$. For $L\to0$, pitch-angle scattering is suppressed and it is found that the Fokker-Planck coefficient tends to zero.
\item For $L$ much larger than the Larmor radius, i.\,e., $L\gg\Rl$, the well-known ``double-hump'' from quasi-linear theory appears. In particular, the Fokker-Planck coefficient becomes symmetric with respect to the negative and positive pitch-angle regimes.
\item For intermediate $L\sim\Rl$, a complicated structure appears in the negative $\mu_0$ regime.
\end{itemize}

%%%%%%%%%%%%%%%%%%%%%%%%%%%%%%%%%%______
%%%%%%%%%%%%%%%%%%%%%%%%%%%%%%%%%%^^^^^^

Furthermore, it can be seen from the results shown above that $\dm$ becomes negative in certain---especially negative---initial pitch-angle cosine regimes. According to Eq.~\eqref{eq:dm_int}, such corresponds to an \emph{anti-correlation} between $\mu(t)$ and $\mu_0$. However, as shown in Fig.~\ref{ab:mu_of_t}, particles with negative initial pitch-angle cosines reverse their parallel motion at some point so that their later pitch-angle cosines become positive, hence the anti-correlation between these two quantities.

Even though a negative diffusion coefficient is puzzling at first glance, it is not unheard of \citep[e.\,g.,][]{chr96:neg,whi08:neg,arg09:neg}. The physical meaning is that of an \emph{inverse} diffusion process, where a concentration gradient is not reduced---as normally the case---but instead particles converge toward some point. For the scenario investigated here, such is supported by the converging magnetic field lines. For a constant turbulence strength (cf. Sec.~\ref{disc}) such will eventually overcome all turbulence-induced scattering and so be responsible for the negative diffusion coefficient.

Specifically, the negative $\dm$ that appeared in the evaluation shown here has to be interpreted in the following way. First, note that for magnetostatic turbulence the pitch-angle derivative,
\be
\dot\mu=\dd t\left(\frac{v\pa}{v}\right)=\frac{1}{v}\,\dd[v\pa]t\propto F\pa,
\ee
corresponds to the force acting on the particles along the field lines. According to Eq.~\eqref{eq:dm_int}, the pitch-angle Fokker-Planck coefficient is defined as the correlation of the parallel force acting on the particle intially and later at later times, which usually is either positive or, in some cases, zero. In addition, it should be noted that in Eq.~\eqref{eq:DmFin} only the turbulent contribution is taken into account. As the turbulence tends to trap particles and to isotropize their velocities, it is to be expected that significant differences to the homogeneous case are to be expected. In particular:

\begin{itemize}
\item Initial pitch-angle cosine $\mu_0\gtrsim-1$: With $\dot\mu(0)>0$ and $\dot\mu(t\leqslant\tau)>0$ with $\tau_c$ the correlation time, it follows that $\dm>0$.
\item Initial pitch-angle cosine $-1<\mu_0<0$:
	\begin{itemize}
	\item For $L\gg\Rl$, the particles turn around slowly so that $\dot\mu$ stays positive for $t\lesssim\tau_c$, yielding $\dm>0$.
	\item For $L\ll\Rl$, the particles are quickly mirrored so that $\dot\mu$ becomes negative due to the directed motion toward larger $z$; accordingly, $\dm<0$.
	\end{itemize}
\item Initial pitch-angle cosine $0<\mu_0\leqslant1$: No turning point is reached and, accordingly, $\dot\mu(t)$ does not on average change its sign so that $\dm>0$.
\end{itemize}

%%%%%%%%%%%%%%%%%%%%%%%%%%%%%%%%%%______
%%%%%%%%%%%%%%%%%%%%%%%%%%%%%%%%%%^^^^^^

It should be noted that a negative Fokker-Planck coefficient is not permitted by the second, approximate expression in Eq.~\eqref{eq:dm_tgk}, which is positive by definition. However, that contradiction is a statement more about the validity of that expression than about the results given here.

%%%%%
\section{Numerical evaluation}\label{sim}

To support the analytical results obtained in the previous sections, a numerical investigation was performed using the \textsc{Padian} code \citep{tau10:pad}. Such codes determine the trajectories of test particles in artificial magnetic turbulence, $\delta\f B$ on top of a background magnetic field, \bo. The turbulence is generated by superposing a number of plane waves with random phase angles and random directions of propagation. The relative weight of each mode is specified by the turbulence power spectrum according to the magnitude---and, for anisotropic turbulence, also the orientation with respect to \bo---of the wave vector. Such simulations are frequently used to investigate: (i) the diffusion of cosmic ray particles; (ii) the random walk of magnetic field lines; or (iii) diffusive or stochastic acceleration processes.

For a sufficiently large number of test particles (here: $2\times10^7$), the Fokker-Planck coefficient for pitch-angle scattering can be obtained \citep[see][]{tau13:pi1} by averaging over all individual contributions. Based on Eq.~\eqref{eq:dm_int}, a time-dependent (``running'') Fokker-Planck coefficient can be derived as
\bs
\begin{eqnarray}
\dm(\mu_0,t)&=&\Bigl\langle\int_0^t\df t'\;\dot\mu(t')\,\dot\mu(0)\Bigr\rangle\\
&=&\bigl\langle\dot\mu(0)\,\De\mu(t)\bigr\rangle
\end{eqnarray}
\es
because $\dot\mu(0)$ is a constant so that only $\dot\mu(t)$ has to be integrated. By sorting all particles according to their original pitch-angle cosine, $\mu_0=\cos\angle\left(\f v(0),\f B(\f r(0))\right)$, the pitch-angle dependence of $\dm$ can be evaluated as required for a comparison with the results in Sec.~\ref{res}.

%%%%%%%%%%%%%%%%%%%%%%%%%%%%%%%%%%______
\begin{figure}[t]
\centering
\includegraphics[bb=95 268 495 558,clip,width=\linewidth]{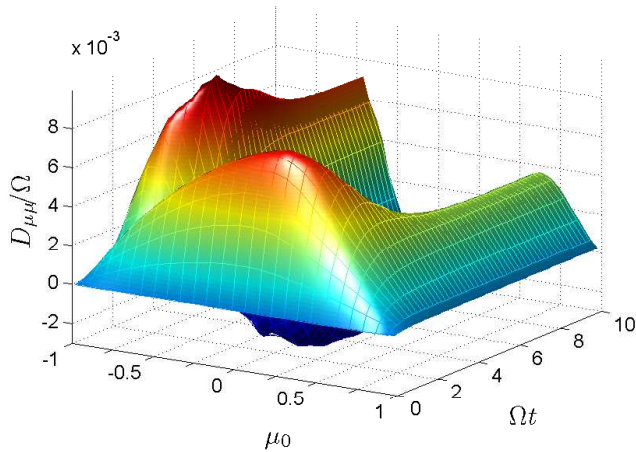}
\caption{(Color online) Running Fokker-Planck coefficient, \dm, as a function of the initial pitch-angle cosine and of the normalized time, $\Omega t$. For the normalized rigidity and the focusing length, values are chosen as $R=1$ and $L=\ell_0$, respectively. Note that the small values for $\dm$ as compared to the previous figures stem from the small turbulence strength that is required to ensure the validity of the quasi-linear limit.}
\label{ab:muCorr}
\end{figure}
%%%%%%%%%%%%%%%%%%%%%%%%%%%%%%%%%%^^^^^^

The result for the running Fokker-Planck coefficient as obtained from the simulation is shown in Fig.~\ref{ab:muCorr}. As seen, there is a considerable variation of $\dm$ as the time increases. In particular, negative values are found only for $\Omega t\gtrsim2$ as predicted by the analytical results. Therefore, it can be concluded that the coherent motion of particles streaming toward weaker magnetic field strengths can lead to a negative $\dm$ especially if for a negative initial pitch-angle cosine, i.\,e., for particles initially moving toward the magnetic bottleneck.

%%%%%
\section{Discussion}\label{disc}

The derivation presented in the previous sections already contains many places where qualifiers and conditions are placed on the results. In addition, several additional points have been left unclear and merit a separate discussion.

\begin{itemize}
\item A discussion as to whether the expression for the Fokker-Planck coefficient $\dm$ given in Eq.~\eqref{eq:dm_tgk} is valid in general has recently been given by \citet{tau13:pi1,tau13:pi2}. It has been shown that the validity is restricted to relatively small times, where the mean-square displacement can still grow without experiencing the ``walls'' resulting from the limits $\mu\in[-1,1]$. For large times, in contrast, the correlation function of the parallel velocity components provides a considerably better agreement due to the generally unlimited value range. Accordingly, the second definition, Eq.~\eqref{eq:dm_tgk} has to be regarded with suspicion for either late times or for strong turbulence that causes significant variations in the particles' pitch angle.

\item It has to be noted that the quasi-linear approximation for the Fourier factor that appears in Eq.~\eqref{eq:fourier} is a dramatic assumption. This corresponds to a drastic simplification of the problem and so must be regarded as not always permitted. A discussion of the problems inherent in the quasi-linear approach and possible non-linear formulations have been given, e.\,g., by \citet{tau06:sta,sha05:soq,tau08:soq,sha09:nli}. Note also that Corrsin's \citeyearpar{cor59:lag,cor59:ind} independence hypothesis has been used to separate the Fourier factor from the correlation function \citep[cf.][]{tau10:rag}.

\item The focusing length has been set to $L=-B_z/(\partial B_z/z)$ which is not strictly correct if $z$ were taken as the Cartesian coordinate. As discussed in \citet[Appendix~A]{tau12:adf}, such a treatment is valid only within a cylinder along the $z$ axis with a limited radius $\sqrt{x^2+y^2}\ll 2L$. A correct treatment with the focusing length, $L$, left constant would require the solution of the equation $1/L=\nabla\cdot(\f B/\abs{B})$ \citep{roe69:int}. Such would result in rather complicated expressions for the magnetic field components and, due to the curved coordinates, would require the re-derivation of the unperturbed particle trajectory.
\end{itemize}

Another interesting point is the turbulence strength in relation to the variable background magnetic field strength. After the expression for the pitch-angle derivative, Eq.~\eqref{eq:DmFull}, was inserted into the original form of the Fokker-Planck coefficient, Eq.~\eqref{eq:dm_int}, the ratio of the turbulent to the background magnetic field was factored out as $\delta B/\abs{\f B}$. Such corresponds to the requirement that said ratio be spatially constant, which is necessary for the perpetuation of the precondition for a quasi-linear treatment that $\delta B\ll\abs{\f B}$. From a physical point of view, however, the opposite case of a spatially constant turbulence strength $\delta B\propto B_0$ cannot be excluded, in which case the ratio of the turbulent to the mean magnetic field would be highly variable. As shown in \citet{tau12:adf}, both cases lead to fundamentally different particle behavior.

As shown in Sec.~\ref{sim}, a precise comparison to the analytical results presented in Sec.~\ref{res} turns out to be difficult for several reasons. Most importantly, the geometry chosen for the analytical derivation is valid only for particles near the central axis, i.\,e., for $\sqrt{x^2+y^2}\ll2L$ \citep[see][]{tau12:adf}. Likewise, the pitch-angle is taken with respect to the $z$ axis whereas, in reality, it refers to the magnetic field orientation. The aforementioned problem of the variable turbulence strength (either absolute or relative) is also likely to influence the outcome of the numerical simulations.

Finally, the applicability of the Fokker-Planck equation, in which the focusing length is a constant parameter, might not always be given. Such is even more true as the Fokker-Planck coefficient is usually taken to be a function of the pitch-angle cosine, $\mu$. Here, in contrast, the initial pitch angle, $\mu_0$, appears as a parameter, which somewhat obscures the use of the Fokker-Planck coefficient. Nevertheless, the transition to the homogeneous case together with the additional physical insight gained by a variable initial pitch angle underlines the validity of our results.

%%%%%
\section{Conclusion}\label{conc}

To describe the small-scale deflections of charged particles in a turbulent plasma in the case of a non-constant background magnetic field, the Fokker-Planck equation with adiabatic focusing is typically used. Here, the Fokker-Planck coefficient of pitch-angle scattering has been derived for the case of a constant adiabatic focusing length. Accordingly, the unperturbed orbit has been derived for a cylindrically symmetric field that fulfills the following conditions: (i) a vanishing azimuthal field component; (ii) a constant focusing length; (iii) the parallel variation depends only weakly on the radial coordinate. Based on the quasi-linear derivation of the Fokker-Planck equation \citep[e.\,g.,][Sec.~12.1]{rs:rays}, the only requirement is that a time scale, $T$, exists so that $t_c\ll T\ll t_F$, where $t_c$ is the correlation time for the turbulent magnetic field and where $t_F$ is the time scale on which the turbulent affects the ensemble-averaged distribution function.

In order to deepen the understanding for the connection between the various parameters entering the Fokker-Planck coefficient and the resulting expressions, additional numerical test-particle simulation might prove useful. However, both the geometry and the time dependence---due to the necessarily finite numerical time in such simulations---seem to be critical issues, as already discussed. Accordingly, obtaining the exact same expressions that have been assumed here for the Fokker-Planck coefficient have turned out to be delicate. The precise comparison with numerical results is therefore left to a future article.

%%%%% Acknowledgements
\acknowledgments

RCT and AD acknowledge fruitful discussions with Fathallah Alouani Bibi and Gary Zank. This work was in part done during RCT's stay at the University of Alabama in Huntsville, which was funded by the \emph{Deut\-sche Aka\-de\-mi\-sche Aus\-tausch\-dienst} (DAAD). AS acknowledges support by the Natural Sciences and Engineering Research Council (NSERC) of Canada. The work of AD is supported by the \emph{Deutsche Forschungsgemeinschaft} (DFG) under grant DO~1505/1-1.

%%%%% Appendix
\appendix

%%%%%
\section{Magnetic Bottle}\label{bottle}

The following derivation is based on the assumption that the magnetic field is predominantly oriented along one of the coordinate axis, which here is the $z$ direction. The classic ``magnetic bottle'' geometry then states that, in addition to the usual Lorentz force causing the gyro-motion, a second force is present along the $z$ axis. It is precisely this force that is used here to describe the adiabatic focusing of particle being scattered in such a magnetic field geometry.

Following \citet{che06:pla}, the two contributions to the Lorentz force are separated using cylindrical coordinates. Therefore, one has
\bs
\be\label{eq:Fz_pre}
F_z=-\frac{q}{c}\,v_\phi B_r,
\ee
causing an additional force \emph{along} the preferred direction of the mean background field and
\be
F\se=\frac{q}{c}\left(v\se\times B_z\right)
\ee
\es
causing normal gyromotion perpendicular to the $z$ axis so that
\bs\label{eq:vxy}
\begin{eqnarray}
v_x(t)&=&v\se\cos(\Ph_0-\Om t)\\
v_y(t)&=&v\se\sin(\Ph_0-\Om t),
\end{eqnarray}
\es
with $v\se$ still to be determined. Note that $\f v\se=v\se\,\hat{\f e}_\phi$ and $\f B_r=B_r\,\hat{\f e}_r$.

Assuming azimuthal symmetry for the magnetic field (i.\,e., $B_\phi=0$), the next step is to use Maxwell's equation $\nabla\cdot\f B=0$ in cylindrical coordinates
\be
\frac{1}{r}\,\pd r\left(rB_r\right)+\pd[B_z]z=0,
\ee
which can be integrated to obtain $rB_r$ as
\be
rB_r=-\int_0^r\df r'\;r'\,\pd[B_z]z.
\ee
Here, $\partial B_z/\partial z$ is assumed to vary only weakly with $r$ (note that, according to Eq.~\eqref{eq:Bz}, $B_z$ does not depend on $x$ and $y$ at all) so that it can be removed from the integral. Then
\be\label{eq:appBr}
rB_r\simeq-\pd[B_z]z\int_0^r\df r\;r=-\frac{1}{2}\,r^2\,\left.\pd[B_z]z\right|_{r\approx0}
\ee
so that the radial magnetic field component can be written
\be\label{eq:Br}
B_r\simeq-\frac{1}{2}\,r\,\pd[B_z]z.
\ee
Note that, for the magnetic field components from Eqs.~\eqref{eq:Bxyz}, Eq.~\eqref{eq:Br} is exactly fulfilled. Alternatively, a different magnetic field structure of course could be assumed, as long as the requirement is fulfilled that $\partial B_z/\partial z$ varies only weakly with $r$ in order to retain the validity of Eq.~\eqref{eq:appBr}.

Now for $r$ use the Larmor radius, $\Rl=\gamma v_\phi/\Om$ and insert Eq.~\eqref{eq:Br} in $F_z$ from Eq.~\eqref{eq:Fz_pre} so that
\be
F_z=\gamma m\dot v\pa\simeq-\frac{\gamma mv\se^2}{2B}\,\pd[B_z]z.
\ee
Note that the relativistic force is given as $F=\df(\gamma mv)/\df t$. If $\f v\perp\f F$ as is the case for the magnetostatic Lorentz force, then $F=\gamma m\dot v$ due to $\gamma=\const$.

%%%%% Bibliography
% \bibliography{../../tautz,../../book,../../article}
% \bibliographystyle{apj}

%%%%% Tables & Figures
% \clearpage

\end{document}